# The existence of stealth corrections in scientific literature - a threat to scientific integrity

*Authors*

René Aquarius[1]

Floris Schoeters[2]

Nick Wise[3]

Alex Glynn[4]

Guillaume Cabanac[5,6]

*Affiliations*

[1] Department of Neurosurgery, Radboud University Medical Center, Nijmegen, The Netherlands.

[2] Radius, Thomas More University of Applied Sciences, Geel, Belgium.

[3] Clare College, University of Cambridge, Cambridge, United Kingdom.

[4] Kornhauser Health Sciences Library, University of Louisville, Louisville (KY), United States of America.

[5] Université Toulouse 3 – Paul Sabatier, IRIT UMR 5505 CNRS, Toulouse, France.

[6] Institut Universitaire de France (IUF), Paris, France

*Author details*

René Aquarius - ORCID: 0000-0002-0968-6884 / E-mail: rene.aquarius@radboudumc.nl

Floris Schoeters - ORCID: 0000-0003-2744-8756 / E-mail: floris.schoeters@thomasmore.be

Nick Wise - ORCID: 0000-0001-7619-7477 / Email: nhw24@cam.ac.uk

Alex Glynn - ORCID: 0000-0002-3027-7276 / E-mail: alex.glynn@louisville.edu

Guillaume Cabanac - ORCID: 0000-0003-3060-6241 / E-mail: guillaume.cabanac@univ-tlse3.fr

*Conflict of interest statement*

All authors declare to have no conflicts of interest.

**Key points**

- Stealth corrections are post-publications changes, without providing any indication that the publication was temporarily or permanently altered.
- We found a total of 131 published articles that were affected by stealth corrections, across different scientific fields.
- Stealth corrections need to end immediately as they threaten scientific integrity.
- We recommend: public logging of all post-publication changes, clear definitions and guidelines, sustained vigilance to register stealth corrections.

**Abstract**

*Introduction*

Thorough maintenance of the scientific record is needed to ensure the trustworthiness of its content. This can be undermined by a stealth correction, which is at least one post-publication change made to a scientific article, without providing a correction note or any other indicator that the publication was temporarily or permanently altered. In this paper we provide several examples of stealth corrections in order to demonstrate that these exist within the scientific literature. As far as we are aware, no documentation of such stealth corrections was previously reported in the scientific literature.

*Methods*

We identified stealth corrections ourselves, or found already reported ones on the public database pubpeer.com or through social media accounts of known science sleuths.

*Results*

In total we report 131 articles that were affected by stealth corrections and were published between 2005 and 2024. These stealth corrections were found among multiple publishers and scientific fields.

*Conclusion and recommendations*

Stealth corrections exist in the scientific literature. This needs to end immediately as it threatens scientific integrity. We recommend the following: 1) Tracking all changes to the published record by all publishers in an open, uniform and transparent manner, preferably by online submission systems that log every change publicly, making stealth corrections impossible; 2) Clear definitions and guidelines on all types of corrections; 3) Support sustained vigilance of the scientific community to publicly register stealth corrections.

## Introduction

One of the cornerstones of publication integrity is the thorough maintenance of the scientific record to ensure the trustworthiness of its content. This includes strict and transparent record keeping when implementing post-publication changes through a clearly visible corrigendum or erratum, which provides details of the changes and the reason why those changes were made (ICMJE, 2024).

However, such record keeping is not always practiced as *stealth changes*, post-publication changes to the scientific literature without any accompanying note, have been observed. One notable kind of a stealth change is a *stealth retraction*: published papers simply disappearing from the website of a journal without a formal retraction notice (Teixeira da Silva, 2016; Teixeira da Silva & Daly, 2024).

Besides stealth retractions, a second problematic type of stealth change exists in the scientific literature: *stealth corrections*. We define a stealth correction as at least one post-publication change made to a scientific article, without providing a correction note or any other indicator that the publication was temporarily or permanently altered.

In this paper we provide examples of stealth corrections. Online blogs have occasionally described such stealth corrections (Bimler, 2021; Schneider, 2020), but a considerable number of stealth corrections appear to have gone unnoticed in the scientific literature.

**Methods**

To document stealth corrections, we applied a pragmatic approach. Stealth corrections are by their nature extremely difficult to track down systematically and it is therefore challenging to provide a comprehensive overview. To document the occurrence of stealth changes, we utilized the following steps:

1. We registered stealth corrections that we had already found;
2. We explored the social media accounts of known science sleuths addressing this type of issues;
3. We contacted science sleuths to ask whether they had encountered stealth corrections;
4. we screened the PubPeer database (Biagioli & Lippman, 2020) to screen for stealth corrections by searching for terms that may refer to stealth corrections: “no erratum”, “no corrigendum”, or “stealth” (repeat the search yourself).

No specific limitations were set regarding the time of occurrence, the journal, or publisher where the stealth corrections took place. Stealth corrections were further categorized into the following types:

1. Changes in author information (addition or removal of authors, changes in author affiliation, etc.);
2. Changes in content (figures, data or text, etc.);
3. Changes in the record of editorial process (editor name, date of submission, acceptance or publication, etc.);
4. changes in additional information (ethics statements, conflicts of interest statements, funding information, etc.).

**Results**

We found a total of 131 published articles that were affected by stealth corrections, across a variety of large and small publishers (Table 1). In most cases, the contents of the articles were changed in secret (Table 2). An overview of all stealth corrections (# 1-131) can be found in Supplementary Table 1, which also contains the accompanying PubPeer post for additional detail. Four articles with stealth corrections eventually received an official correction notice and 17 articles were reverted to their original version (Supplementary Table 1). Nine expressions of concern were published, and 11 articles were eventually retracted (Supplementary Table 1).

Table 1: number of published articles affected by stealth corrections per publisher.

| Publisher | Articles with stealth corrections (n) |
|---|---|
| BAKIS Productions LTD | 75 |
| MDPI | 25 |
| Elsevier | 22 |
| IEEE | 2 |
| SpringerNature | 2 |
| ACS Publications | 1 |
| Academy Of Traditional Chinese Medicine | 1 |
| American Association for the Advancement of Science | 1 |
| IntechOpen | 1 |
| Tech Science Press | 1 |

Table 2: number of published articles affected by stealth corrections per correction type.

| Type of correction | Articles with stealth corrections (n) |
|---|---|
| Changes in content | 92 |
| Changes in the record of editorial process | 27 |
| Changes in author information | 9 |
| Changes in additional information | 3 |

**Discussion**

The stealth corrections presented in this paper demonstrate a fundamental and mostly ignored problem in the scientific literature. Correct documentation and transparency are of the utmost importance to uphold scientific integrity and the trustworthiness of science. Post-publication changes need to be clear in order for readers to understand if, and why, changes have been made. However, little attention is paid to post-publication alterations. For example, the Committee on Publication Ethics (COPE) provides clear flow charts indicating that a published correction is needed in case of: post-publication plagiarism (# 20) (COPE, 2021b), post-publication author removal (# 2-9) (COPE, 2021a) or inappropriate image manipulation in a published article (# 10-11, 18-19, 21-101) (COPE, 2024). However, no specific flow charts exist on post-publication changes in author affiliations (# 1-2), consent or ethics statements (# 129-130), handling editor (# 102-103), or allocation to journal issue (# 104-128). Post-publication amendments that are made silently, without a visible correction note, will give rise to questions regarding the ethics and integrity of the specific journal, editors and publisher, and might undermine the validity of the published literature as a whole. We therefore strongly encourage that every post-publication change is publicly logged in the interest of transparency.

It is particularly concerning when publications with serious problems such as image duplication or data duplication receive stealth corrections. Future readers are likely to be unaware that such problems were ever present, which might lead to less critical assessment of these publications. Readers of our findings might even get the impression that editors or publishers are actively trying to hide problems, such as in one case where the journal's Editor-in-chief was listed as an author of the corrected paper (# 19) or the cases where the publisher removed an article authored by a special from that special issue, possibly in order to meet the criteria of the Directory of Open Access Journals (DOAJ) (# 104-128).

We need clear definitions of the term "correction". Situations can get complex very quickly as publishers often post preliminary or pre-publication versions of papers while they are undergoing type setting or even pre-prints on their own hosting platform. What is the precise moment of publication in these cases? If anything changes during these stages, will these changes count as corrections? Should undocumented, fundamental changes such as the replacing of images or results during this phase of the publishing process still be regarded as stealth corrections or not? By logging every change made to an article that can be read and cited, no matter the stage

Another complicating factor is that publishers might use internal guidelines on how to handle corrections. These guidelines might leave room for interpretation by its reader, as in the case with Elsevier: "*If […] new material should replace the original content of the accepted article, the editor may consider the publication of an erratum or a corrigendum*" (Elsevier, 2024). Guidelines might also lack the information needed for editors to make a decision on the topic, as is the case with MDPI's guidelines that do not describe what to do when removing articles from a special issue (MDPI, 2024). The abovementioned COPE flow charts might help to circumvent these problems and could fulfill publisher-independent guidance on how to handle post-publication corrections. However, additional flow charts are needed to adequately handle all types of post-publication corrections.

Stealth corrections add yet another threat to the integrity of scientific publishing, besides previously described phenomena such as paper mills (Bishop & Abalkina, 2023; Candal-Pedreira et al., 2022; Else & Van Noorden, 2021), fake peer review (Oviedo-García, 2024) and abuse of special issues (Mills et al., 2024). Authors are often at the heart of these problems, fueled by the harsh publish-or-perish culture of academia (Al-Adawi et al., 2016; Vasconez-Gonzalez et al., 2024). However, adequate handling of corrections is exclusively the responsibility of publishers, journals and editorial boards. Academic editors are often seen as the safeguards of the scientific literature and its integrity (Desai & Shortell, 2011; Marusic, 2010), but in the case of stealth corrections, this integrity can come into question.

Our results demonstrate that the scientific community needs to stay vigilant in order to detect and report stealth corrections. Reporting should preferably take place on a public platform such as PubPeer to provide the necessary transparency and long-term documentation.

In order to add new cases and monitor the cases we have described in this article, we have created an online spreadsheet listing stealth corrections (find it here). We invite interested readers to forward any PubPeer link detailing stealth corrections to us, so we can add them to the spreadsheet.

**Conclusion and recommendations**

Stealth corrections exist in the scientific literature. This needs to end immediately as it threatens scientific integrity. We recommend the following:

1) Tracking of all changes to the published record by all publishers in an open, uniform and transparent manner, preferably by online submission systems that log every change publicly, making stealth corrections impossible;

2) Clear definitions and guidelines on all types of corrections;

3) Support sustained vigilance of the scientific community to publicly register stealth corrections.

**Acknowledgements**

We thank Dorothy Bishop, Jana Christopher and David Bimler for proof-reading our manuscript and giving valuable feedback. We thank all (anonymous) science sleuths who have found and reported stealth corrections, your work is much appreciated.

Supplemental Table 1: Stealth corrections collected during this work grouped per type of stealth correction. *Last assessed between August 14-26, 2024

| # | DOI or PMID | Journal | Publisher | Problem | Pubpeer link | Resolved? * |
|---|---|---|---|---|---|---|
| | | | *Changes in author information* | | | |
| 1 | https://doi.org/10.1016/j.comptc.2024.114495 | Computational & theoretical chemistry | Elsevier | An author affiliation was removed without notice as described by *Nick Wise* and *Francois-Xavier Coudert* in March 2024. | https://pubpeer.com/publications/EB0331DBF5F262125FEC913C1F48A8#2 | No |
| 2 | https://doi.org/10.1016/j.mtcomm.2024.108363 | Materials Today Communications | Elsevier | An author affiliation was removed without notice as described by *François-Xavier Coudert* in February 2024. | https://pubpeer.com/publications/697C3F6F5FFEFF680E507DEDF86A52#3 | No |
| 3 | https://doi.org/10.1016/bs.pbr.2020.09.011 | Progress in brain research (book) volume 257 - Chapter III | Elsevier | An author was removed from the author list without notice as described by *Bythinella padiraci* in November 2023. | https://pubpeer.com/publications/405E69FA1DCB69B8790FB8FCB09929#3 | No |
| 4 | https://doi.org/10.1016/bs.pbr.2020.09.010 | Progress in brain research (book) volume 258 - Chapter IV | Elsevier | An author was removed from the author list without notice as described by *Bythinella padiraci* in November 2023. | https://pubpeer.com/publications/3D64DDE95DB292E7D2D798722D99C3#3 | No |
| 5 | https://doi.org/10.1016/bs.pbr.2020.09.009 | Progress in brain research (book) volume 258 - Chapter V | Elsevier | An author was removed from the author list without notice as described by *Bythinella padiraci* in November 2023 | https://pubpeer.com/publications/C54D1DF4F4B753AE49600115C39878#4 | No |
| 6 | https://doi.org/10.1016/bs.pbr.2020.09.004 | Progress in brain research (book) volume 258 - Chapter VI | Elsevier | An author was removed from the author list without notice as described by *Bythinella padiraci* in November 2023. | https://pubpeer.com/publications/4CF02DEE9E54A30B0FE9974654BF66#4 | No |
| 7 | https://doi.org/10.1016/bs.pbr.2021.06.003 | Progress in brain research (book) volume 266 – Chapter I | Elsevier | An author was removed from the author list without notice as described by *Bythinella padiraci* in November 2023. | https://pubpeer.com/publications/CD1F2980DA29D2F144910B21A04E56#4 | No |
| 8 | https://doi.org/10.1016/bs.pbr.2021.06.008 | Progress in brain research (book) volume 266 - Chapter IV | Elsevier | An author was removed from the author list without notice as described by *Bythinella padiraci* in November 2023. | https://pubpeer.com/publications/7D493EE04E6E90EE6A53001C1E0D8F#4 | No |
| 9 | https://doi.org/10.1016/bs.pbr.2021.06.014 | Progress in brain research (book) volume 266 - Chapter VIII | Elsevier | An author was removed from the author list without notice as described by *Desmococcus antarctica* November 2023. | https://pubpeer.com/publications/D9683206D93312C22E3B73AB176468#7 | No |
| | | | *Changes in content* | | | |
| 10 | https://doi.org/10.1016/j.neulet.2016.04.018 | Neuroscience Letters | Elsevier | An image with duplication issues was replaced without notice as described by *Rene Aquarius* in April 2024. | https://pubpeer.com/publications/8A0A30197C521C60F66C30EE04ED28#3 | Yes, after mailing the ethics department about it, the issue was resolved. The stealth correction was present for about 3 months. |
| 11 | https://doi.org/10.1016/j.algal.2024.103400 | Algal Research | Elsevier | An image and data were corrected without notice as described by *Hyponephele interposita* in July 2024. | https://pubpeer.com/publications/32FAC1A793369B3991EA13E47E3A7B#4 | No |
| 12 | https://doi.org/10.1016/j.inffus.2023.101847 | Information Fusion | Elsevier | The phrase "Regenerate response", accidentally copied from the ChatGPT interface, was removed; a note about ChatGPT was added to the acknowledgements as described by *Guillaume Cabanac* in March 2024. | https://pubpeer.com/publications/6A8918534C304C6BD58F51DC126838#3 | Yes, a correction was issued in May 2024. The stealth correction was present for at least 1 month. |
| 13 | https://doi.org/10.1016/j.tox.2023.153629 | Toxicology | Elsevier | The phrase "Regenerate response", likely copied from the ChatGPT interface, was removed (from the pre-proof version), as described by *Alex Glynn* in August 2024. | https://pubpeer.com/publications/8C16BAC6189B02EAE9192E7D5D28C2#3 | No |
| 14 | https://doi.org/10.1016/j.nlp.2023.100027 | Natural Language Processing Journal | Elsevier | The phrase "Regenerate response", likely copied from the ChatGPT interface, was removed between July 27 and September 19, 2023, as described by *Guillaume Cabanac* in November 2023. | https://pubpeer.com/publications/82AE520EA54FC025369F9CD6E17C68#4 | No |
| 15 | https://doi.org/10.1016/j.jormas.2022.04.011 | Journal of Stomatology Oral and Maxillofacial Surgery | Elsevier | The name of the study's institution, previously redacted to 'XXX', was restored without a formal notice, as described by *Guillaume Cabanac* in March 2024. | https://pubpeer.com/publications/795E241C8075BAE2229AD880C7ABE3#2 | No |
| 16 | https://doi.org/10.1016/j.aimed.2022.03.001 | Advances in Integrative Medicine | Elsevier | The name of the study's institution, previously redacted to 'XXX', was restored without a formal notice, as described by *Guillaume Cabanac* in February 2024. | https://pubpeer.com/publications/4D7D4EFCA4C1B42258F6EC003D9ED5#2 | No |
| 17 | https://doi.org/10.1007/s00784-022-04384-2 | Clinical Oral Investigations | SpringerNature | The name of the study's institution, previously redacted to 'XXX', was restored without a formal notice, as described by Guillaume Cabanac in February 2024. | https://pubpeer.com/publications/E947C970662AF57D43F4EC896487C6#2 | No |
| 18 | https://doi.org/10.1021/acs.nanolett.3c02933 | Nano Letters | ACS Publications | An image with duplication issues was replaced without notice as described by *Pseudanarta actura* in April 2024. | https://pubpeer.com/publications/F7AC72D56DA9A97AA8EAE1CE8A2B08#3 | No |
| 19 | https://doi.org/10.1007/s12035-022-02937-w | Molecular Neurobiology | SpringerNature | An image with duplication issues was replaced without notice as described by *Dysdera arabisenen* in August 2023. The editor in chief of the journal at that time was also co-author of this paper. | https://pubpeer.com/publications/924331EB2AFA8615F307EE640D0F55#3 | Yes, a correction was issued in February 2024 and the issue was resolved. The stealth correction was present for at least 7 months. |
| 20 | https://doi.org/10.5772/20640 | Advances in Nanocomposite Technology | IntechOpen | Text has been copied verbatim from a 2009 paper published in Physical Review B. The text was replaced and a reference to that paper was added without notice as described by *Desulfofrigus oceanense* in January 2024. | https://pubpeer.com/publications/39EC348023D0F65D48DDA33A2DB41A#2 | Yes, a correction was issued in February 2024 and the issue was resolved. The stealth correction was present for at least 1 month. |
| 21 | https://doi.org/10.1126/sciadv.abc3099 | Science advances | American Association for the Advancement of Science | An image with duplication issues was replaced without notice as described by *Tabuina rufa* in April 2021 | https://pubpeer.com/publications/6FFF7386A0DE678EE5886D17CA3480#2 | No |
| 22 | https://doi.org/10.3727/096504017x14934860122864 | Oncology Research | Tech Science Press | An image with duplication issues was replaced without notice as described by *Actinopolyspora biskrensis* in August 2023. | https://pubpeer.com/publications/97477B8D75C46117EA8526736D430A#4 | No |
| 23 | BAKIS Productions LTD | Journal of B.U.ON. | BAKIS Productions LTD | An image with duplication issues was replaced without notice as described by *Hoya camphorifolia* in February 2021. | https://pubpeer.com/publications/07F23F107DD25ED59BCC83D2007794#3 | No |
| 24 | https://pubmed.ncbi.nlm.nih.gov/32521863 | Journal of B.U.ON. | BAKIS Productions LTD | An image with duplication issues was replaced without notice as described by *Hoya camphorifolia* in February 2021. | https://pubpeer.com/publications/08916C0F17BCC06FA9EA164DD91B12#2 | No |
| 25 | https://pubmed.ncbi.nlm.nih.gov/33099954 | Journal of B.U.ON. | BAKIS Productions LTD | An image with duplication issues was replaced without notice as described by Hoya *camphorifolia* in January 2021. | https://pubpeer.com/publications/08A5CA9E7A9F6E2CCB86897C0B45E6#2 | No |
| 26 | https://pubmed.ncbi.nlm.nih.gov/33277851 | Journal of B.U.ON. | BAKIS Productions LTD | An image with duplication issues was replaced without notice as described by *Hoya camphorifolia* in January 2021. | https://pubpeer.com/publications/136066E0869A9722DAF0424D0D179A#2 | No, the publication was retracted in July/August of 2021. |
| 27 | https://pubmed.ncbi.nlm.nih.gov/32862589 | Journal of B.U.ON. | BAKIS Productions LTD | An image with duplication issues was replaced without notice as described by *Hoya camphorifolia* in February 2021. | https://pubpeer.com/publications/16CB9E1DEEA94BED25A3266502E8C1#3 | No |
| 28 | https://pubmed.ncbi.nlm.nih.gov/32277651 | Journal of B.U.ON. | BAKIS Productions LTD | An image with duplication issues was replaced without notice as described by *Hoya camphorifolia* in February 2021. | https://pubpeer.com/publications/1791897C706C0C34EE64173849E1B6#4 | No, an expression of concern was issued in March/April 2021. The publication was retracted in November/December 2021. |
| 29 | https://pubmed.ncbi.nlm.nih.gov/33277839 | Journal of B.U.ON. | BAKIS Productions LTD | An image with duplication issues was replaced without notice as described by *Hoya camphorifolia* in January 2021. | https://pubpeer.com/publications/19807296DAAD348CC8AEF07DA73C33#2 | No |
| 30 | https://pubmed.ncbi.nlm.nih.gov/32277631 | Journal of B.U.ON. | BAKIS Productions LTD | An image with duplication issues was replaced without notice as described by *Hoya camphorifolia* in February 2021. | https://pubpeer.com/publications/1D87C1AA4B3015F3CCD50FFCAE8039#2 | No |
| 31 | https://pubmed.ncbi.nlm.nih.gov/32862599 | Journal of B.U.ON. | BAKIS Productions LTD | An image with duplication issues was replaced without notice as described by *Hoya camphorifolia* in February 2021. | https://pubpeer.com/publications/21FB7BEE0475B0CC825D239BAAC222#2 | No |
| 32 | https://pubmed.ncbi.nlm.nih.gov/26537082 | Journal of B.U.ON. | BAKIS Productions LTD | An image with duplication issues was replaced without notice as described by *Hoya camphorifolia* in February 2021. | https://pubpeer.com/publications/2537DE565C3AEB53C656CE8BA9BED4#6 | No |
| 33 | https://pubmed.ncbi.nlm.nih.gov/30941986 | Journal of B.U.ON. | BAKIS Productions LTD | An image with duplication issues was replaced without notice as described by Hoya camphorifolia in January 2021. | https://pubpeer.com/publications/2588FF8B58030ECACA7AE7397C555E#4 | No |
| 34 | https://pubmed.ncbi.nlm.nih.gov/30003747 | Journal of B.U.ON. | BAKIS Productions LTD | An image with duplication issues was replaced without notice as described by *Hoya camphorifolia* in February 2021. | https://pubpeer.com/publications/25EE82E3DC470A25C69C70FE4AFFA4#4 | No |
| 35 | https://pubmed.ncbi.nlm.nih.gov/31786876 | Journal of B.U.ON. | BAKIS Productions LTD | An image with duplication issues was replaced without notice as described by *Hoya camphorifolia* in January 2021. | https://pubpeer.com/publications/26D9696913A57A05921C4959A01860#7 | No, the publication was retracted in July/August 2021. |
| 36 | https://pubmed.ncbi.nlm.nih.gov/30003744 | Journal of B.U.ON. | BAKIS Productions LTD | An image with duplication issues was replaced without notice as described by *Hoya camphorifolia* in February 2021. | https://pubpeer.com/publications/2707D62D903E6D08401D4FBEC99AD7#2 | No |
| 37 | https://pubmed.ncbi.nlm.nih.gov/29745075 | Journal of B.U.ON. | BAKIS Productions LTD | An image with duplication issues was replaced without notice as described by *Hoya camphorifolia* in February 2021. | https://pubpeer.com/publications/303BA09E0D24821C82782F276B4F11#3 | No |
| 38 | https://pubmed.ncbi.nlm.nih.gov/32277661 | Journal of B.U.ON. | BAKIS Productions LTD | An image with duplication issues was replaced without notice as described by *Hoya camphorifolia* in January 2021. | https://pubpeer.com/publications/31A581FCD14FA9493E57075A8EFEB8#3 | No, the publication has been reverted to its original form. This has been documented by *Hoya camphorifolia* in September 2021. The stealth correction was present for 9 months at most. An expression of concern has been issued in May/June 2021. |
| 39 | https://pubmed.ncbi.nlm.nih.gov/32521856 | Journal of B.U.ON. | BAKIS Productions LTD | An image with duplication issues was replaced without notice as described by *Hoya camphorifolia* in January 2021. | https://pubpeer.com/publications/3413E65249A90A68C916F7256188AE#2 | No |
| 40 | https://pubmed.ncbi.nlm.nih.gov/30941961 | Journal of B.U.ON. | BAKIS Productions LTD | An image with duplication issues was replaced without notice as described by *Hoya camphorifolia* in January 2021. | https://pubpeer.com/publications/392F1600DFD18541F13613DEE7E07D#7 | No, the publication was retracted in November/December of 2021. |
| 41 | https://pubmed.ncbi.nlm.nih.gov/31128032 | Journal of B.U.ON. | BAKIS Productions LTD | An image with duplication issues was replaced without notice as described by *Hoya camphorifolia* in January 2021. | https://pubpeer.com/publications/3A3BF074DCBA381EDA5453454CD15C#3 | No |
| 42 | https://pubmed.ncbi.nlm.nih.gov/26854457 | Journal of B.U.ON. | BAKIS Productions LTD | An image with duplication issues was replaced without notice as described by *Hoya camphorifolia* in February 2021. | https://pubpeer.com/publications/3C8EA70081485AF27340183A6CF7EF#4 | No, the publication has been reverted to its original form. This has been documented by *Hoya camphorifolia* in February 2022. The stealth correction was present for 12 months at most. |
| 43 | https://pubmed.ncbi.nlm.nih.gov/30003745 | Journal of B.U.ON. | BAKIS Productions LTD | An image with duplication issues was replaced without notice as described by *Hoya camphorifolia* in February 2021. | https://pubpeer.com/publications/3DBF57A1BEC46993276E23461201D0#3 | No |
| 44 | https://pubmed.ncbi.nlm.nih.gov/26854445 | Journal of B.U.ON. | BAKIS Productions LTD | An image with duplication issues was replaced without notice as described by *Hoya camphorifolia* in February 2021. | https://pubpeer.com/publications/3E79DCFC48E5B748F8C71D92D31B34#4 | No, the publication has been reverted to its original form. This has been documented by *Hoya camphorifolia* in June 2022. The stealth correction was present for 17 months at most. |
| 45 | https://pubmed.ncbi.nlm.nih.gov/32521859 | Journal of B.U.ON. | BAKIS Productions LTD | An image with duplication issues was replaced without notice as described by *Hoya camphorifolia* in January 2021. | https://pubpeer.com/publications/4362F3CD24CF9D1B86A0089BF278F6#2 | No, the publication was retracted in November/December 2021. |
| 46 | https://pubmed.ncbi.nlm.nih.gov/32277664 | Journal of B.U.ON. | BAKIS Productions LTD | An image with duplication issues was replaced without notice as described by *Hoya camphorifolia* in February 2021. | https://pubpeer.com/publications/49DACDAFF8725AA5ADB0193660B737#4 | No, the publication has been reverted to its original form. This |
| | | | | | | has been documented by *Hoya camphorifolia* in September 2021. The stealth correction was present for 8 months at most. An expression of concern has been issued in May/June 2021. |
| 47 | https://pubmed.ncbi.nlm.nih.gov/32277642 | Journal of B.U.ON. | BAKIS Productions LTD | An image with duplication issues was replaced without notice as described by *Hoya camphorifolia* in January 2021. | https://pubpeer.com/publications/5359FFF5D1BB9974ABC28E474A31F6#2 | No, an expression of concern was issued in March/April 2021. |
| 48 | https://pubmed.ncbi.nlm.nih.gov/32862572 | Journal of B.U.ON. | BAKIS Productions LTD | An image with duplication issues was replaced without notice as described by *Hoya camphorifolia* in January 2021. | https://pubpeer.com/publications/546F3264F06F122264355278AE8C4A#4 | No |
| 49 | https://pubmed.ncbi.nlm.nih.gov/32521844 | Journal of B.U.ON. | BAKIS Productions LTD | An image with duplication issues was replaced without notice as described by *Hoya camphorifolia* in January 2021. | https://pubpeer.com/publications/5A8F97D8A56CAA0714C5BF573F4E29#2 | No |
| 50 | https://pubmed.ncbi.nlm.nih.gov/32521905 | Journal of B.U.ON. | BAKIS Productions LTD | An image with duplication issues was replaced without notice as described by *Hoya camphorifolia* in February 2021. | https://pubpeer.com/publications/625665098ACC61B9B0A741575F0FA6#3 | No |
| 51 | https://pubmed.ncbi.nlm.nih.gov/30003743 | Journal of B.U.ON. | BAKIS Productions LTD | An image with duplication issues was replaced without notice as described by *Hoya camphorifolia* in February 2021. | https://pubpeer.com/publications/67BD6877FAED9BFA5747B4DAB3FB0A#2 | No |
| 52 | https://pubmed.ncbi.nlm.nih.gov/27061538 | Journal of B.U.ON. | BAKIS Productions LTD | An image with duplication issues was replaced without notice as described by *Hoya camphorifolia* in February 2021. | https://pubpeer.com/publications/683F667BD4F84594995591D0E33237#8 | No, the publication was retracted in November/December of 2021. |
| 53 | https://pubmed.ncbi.nlm.nih.gov/30941982 | Journal of B.U.ON. | BAKIS Productions LTD | An image with duplication issues was replaced without notice as described by *Hoya camphorifolia* in January 2021. | https://pubpeer.com/publications/6B8AB71A32DBE137501B0BB8E066E2#2 | No |
| 54 | https://pubmed.ncbi.nlm.nih.gov/31424660 | Journal of B.U.ON. | BAKIS Productions LTD | An image with duplication issues was replaced without notice as described by *Hoya camphorifolia* in January 2021. | https://pubpeer.com/publications/6C7046FBA99A9C1103B3CB7FEAB9FF#2 | No |
| 55 | https://pubmed.ncbi.nlm.nih.gov/31786886 | Journal of B.U.ON. | BAKIS Productions LTD | An image with duplication issues was replaced without notice as described by *Hoya camphorifolia* in January 2021. | https://pubpeer.com/publications/745E86F421CF39E07DA24667AC4864#7 | No, the publication has been reverted to its original form. This has been documented by *Hoya* |
| | | | | | | *camphorifolia* in June 2022. The stealth correction was present for 18 months at most. |
| 56 | https://pubmed.ncbi.nlm.nih.gov/33277868 | Journal of B.U.ON. | BAKIS Productions LTD | An image with duplication issues was replaced without notice as described by *Hoya camphorifolia* in January 2021. | https://pubpeer.com/publications/771875DF895156FDB92E8650F178E9#3 | No, the publication was retracted in July/August 2021. |
| 57 | https://pubmed.ncbi.nlm.nih.gov/32277640 | Journal of B.U.ON. | BAKIS Productions LTD | An image with duplication issues was replaced without notice as described by *Hoya camphorifolia* in February 2021. | https://pubpeer.com/publications/7A203DED7A18CB6F476883A97F3A3D#3 | No, an expression of concern was issued in March/April 2021. |
| 58 | https://pubmed.ncbi.nlm.nih.gov/30358211 | Journal of B.U.ON. | BAKIS Productions LTD | An image with duplication issues was replaced without notice as described by *Hoya camphorifolia* in February 2021. | https://pubpeer.com/publications/7B9B3BCA7DCCBD951271D6EF06D913#2 | No |
| 59 | https://pubmed.ncbi.nlm.nih.gov/33277865 | Journal of B.U.ON. | BAKIS Productions LTD | An image with duplication issues was replaced without notice as described by *Hoya camphorifolia* in January 2021. | https://pubpeer.com/publications/84EAE1029A59A777FBBB84A90DABE8#5 | No, the publication has been reverted to its original form. This has been documented by *Hoya camphorifolia* in September 2021. The stealth correction was present for 9 months at most. The publication was retracted in July/August 2021. |
| 60 | https://pubmed.ncbi.nlm.nih.gov/32277674 | Journal of B.U.ON. | BAKIS Productions LTD | An image with duplication issues was replaced without notice as described by *Hoya camphorifolia* in January 2021. | https://pubpeer.com/publications/8D5226574DADD0FE551F97E8ECEA5C#2 | No, the publication has been reverted to its original form. This has been documented by *Hoya camphorifolia* in September 2021. The stealth correction was present for 9 months at most. An expression of concern has been issued in May/June 2021. |
| 61 | https://pubmed.ncbi.nlm.nih.gov/28365941 | Journal of B.U.ON. | BAKIS Productions LTD | An image with duplication issues was replaced without notice as described by *Hoya camphorifolia* in February 2021. | https://pubpeer.com/publications/900A0653DD243BC319D578E757F2ED#2 | No |
| 62 | https://pubmed.ncbi.nlm.nih.gov/32521882 | Journal of B.U.ON. | BAKIS Productions LTD | An image with duplication issues was replaced without notice as described by *Hoya camphorifolia* in January 2021. | https://pubpeer.com/publications/939CB49AEF9C729E833F16AAF6D393#2 | No |
| 63 | https://pubmed.ncbi.nlm.nih.gov/29332353 | Journal of B.U.ON. | BAKIS Productions LTD | An image with duplication issues was replaced without notice as described by *Hoya camphorifolia* in January 2021. | https://pubpeer.com/publications/97559A8CC9736A3BB609B32DBB251F#9 | No |
| 64 | https://pubmed.ncbi.nlm.nih.gov/31424680 | Journal of B.U.ON. | BAKIS Productions LTD | An image with duplication issues was replaced without notice as described by *Hoya camphorifolia* in January 2021. | https://pubpeer.com/publications/982A0227D75C92C952E7D1930D9916#2 | No |
| 65 | https://pubmed.ncbi.nlm.nih.gov/27569083 | Journal of B.U.ON. | BAKIS Productions LTD | An image with duplication issues was replaced without notice as described by *Hoya camphorifolia* in January 2021. | https://pubpeer.com/publications/9EAA55137909493E567E4A8EAC479E#4 | No |
| 66 | https://pubmed.ncbi.nlm.nih.gov/29552787 | Journal of B.U.ON. | BAKIS Productions LTD | An image with duplication issues was replaced without notice as described by *Hoya camphorifolia* in February 2021. | https://pubpeer.com/publications/9ECEE6BAEB97D3E795CB4AD919F59E#4 | No |
| 67 | https://pubmed.ncbi.nlm.nih.gov/31646812 | Journal of B.U.ON. | BAKIS Productions LTD | An image with duplication issues was replaced without notice as described by *Hoya camphorifolia* in January 2021. | https://pubpeer.com/publications/A5ADDAAA1995F9A39C5F749FDCC846#3 | No, the publication has been reverted to its original form. This has been documented by *Hoya camphorifolia* in June 2022. The stealth correction was present for 18 months at most. |
| 68 | https://pubmed.ncbi.nlm.nih.gov/32277668 | Journal of B.U.ON. | BAKIS Productions LTD | An image with duplication issues was replaced without notice as described by *Hoya camphorifolia* in January 2021. | https://pubpeer.com/publications/A72A475490A7FD970D4106DF74D7AF#3 | No, the publication has been reverted to its original form. This has been documented by *Hoya camphorifolia* in September 2021. The stealth correction was present for 9 months at most. An expression of concern |
| | | | | | | has been issued in May/June 2021. |
| 69 | https://pubmed.ncbi.nlm.nih.gov/33277849 | Journal of B.U.ON. | BAKIS Productions LTD | An image with duplication issues was replaced without notice as described by *Hoya camphorifolia* in January 2021. | https://pubpeer.com/publications/A8564995D7FE3A488036969678E34F#4 | No |
| 70 | https://pubmed.ncbi.nlm.nih.gov/26537084 | Journal of B.U.ON. | BAKIS Productions LTD | An image with duplication issues was replaced without notice as described by *Hoya camphorifolia* in January 2021. | https://pubpeer.com/publications/AB512DFE29D9365C0DC33AF9EE55C1#8 | No |
| 71 | https://pubmed.ncbi.nlm.nih.gov/30941981 | Journal of B.U.ON. | BAKIS Productions LTD | An image with duplication issues was replaced without notice as described by *Hoya camphorifolia* in January 2021. | https://pubpeer.com/publications/ACB980C6A820AD95177666FBEC20D1#6 | No |
| 72 | https://pubmed.ncbi.nlm.nih.gov/33099951 | Journal of B.U.ON. | BAKIS Productions LTD | An image with duplication issues was replaced without notice as described by *Hoya camphorifolia* in January 2021. | https://pubpeer.com/publications/AD04EFE6B302EA3A5D614A9C12DCE4#2 | No |
| 73 | https://pubmed.ncbi.nlm.nih.gov/27685910 | Journal of B.U.ON. | BAKIS Productions LTD | An image with duplication issues was replaced without notice as described by *Hoya camphorifolia* in January 2021. | https://pubpeer.com/publications/B4C502A30F93F28D8AEFB4A4EC9156#3 | No |
| 74 | https://pubmed.ncbi.nlm.nih.gov/31424647 | Journal of B.U.ON. | BAKIS Productions LTD | An image with duplication issues was replaced without notice as described by *Hoya camphorifolia* in February 2021. | https://pubpeer.com/publications/B6426DECD29B7E73D45B1E46C6634F#2 | No, the publication has been reverted to its original form. This has been documented by *Hoya camphorifolia* in July 2022. The stealth correction was present for 18 months at most. |
| 75 | https://pubmed.ncbi.nlm.nih.gov/33099940 | Journal of B.U.ON. | BAKIS Productions LTD | An image with duplication issues was replaced without notice as described by *Hoya camphorifolia* in February 2021. | https://pubpeer.com/publications/BA4E2ED6394601929E3DE87B3BDA4B#2 | No |
| 76 | https://pubmed.ncbi.nlm.nih.gov/32521851 | Journal of B.U.ON. | BAKIS Productions LTD | An image with duplication issues was replaced without notice as described by *Hoya camphorifolia* in January 2021. | https://pubpeer.com/publications/BC2CC3E2C98B11BDB399E98E5D92DC#3 | The publication has been reverted to its original form. This has been documented by *Hoya camphorifolia* in September 2021. The stealth correction was present for 9 months at most. An |
| | | | | | | expression of concern has been issued in May/June 2021. |
| 77 | https://pubmed.ncbi.nlm.nih.gov/31128007 | Journal of B.U.ON. | BAKIS Productions LTD | An image with duplication issues was replaced without notice as described by *Hoya camphorifolia* in January 2021. | https://pubpeer.com/publications/BF863C7EA33D348B234D4D8D588323#7 | No |
| 78 | https://pubmed.ncbi.nlm.nih.gov/33277838 | Journal of B.U.ON. | BAKIS Productions LTD | An image with duplication issues was replaced without notice as described by *Hoya camphorifolia* in January 2021. | https://pubpeer.com/publications/C37F2B9350FB566C1939A774C7FD41#3 | No |
| 79 | https://pubmed.ncbi.nlm.nih.gov/30941973 | Journal of B.U.ON. | BAKIS Productions LTD | An image with duplication issues was replaced without notice as described by *Hoya camphorifolia* in January 2021. | https://pubpeer.com/publications/C3B7F47FE09EF49A06240E8C5EE43F#3 | The publication was retracted in November/December of 2021 (https://jbuon.com/archive/26-6-2721.pdf). |
| 80 | https://pubmed.ncbi.nlm.nih.gov/32862611 | Journal of B.U.ON. | BAKIS Productions LTD | An image with duplication issues was replaced without notice as described by *Hoya camphorifolia* in January 2021. | https://pubpeer.com/publications/C903D30891B8C23C6CE647F5949FEB#3 | No |
| 81 | https://pubmed.ncbi.nlm.nih.gov/32862591 | Journal of B.U.ON. | BAKIS Productions LTD | An image with duplication issues was replaced without notice as described by *Hoya camphorifolia* in January 2021. | https://pubpeer.com/publications/CEE86FB2BDA77B249A8FB67F2AF487#5 | No |
| 82 | https://pubmed.ncbi.nlm.nih.gov/31786887 | Journal of B.U.ON. | BAKIS Productions LTD | An image with duplication issues was replaced without notice as described by *Hoya camphorifolia* in January 2021. | https://pubpeer.com/publications/D118A33567EEA617C94C86C6C6732D#2 | No |
| 83 | https://pubmed.ncbi.nlm.nih.gov/32862573 | Journal of B.U.ON. | BAKIS Productions LTD | An image with duplication issues was replaced without notice as described by *Hoya camphorifolia* in January 2021. | https://pubpeer.com/publications/D3D144A9E5A7DB80A4274A0FE9B47C#5 | No |
| 84 | https://pubmed.ncbi.nlm.nih.gov/31646794 | Journal of B.U.ON. | BAKIS Productions LTD | An image with duplication issues was replaced without notice as described by *Hoya camphorifolia* in January 2021. | https://pubpeer.com/publications/D4D6B3641B553BAF1A44831ED5F753#2 | No, the publication was retracted in November/December of 2021. |
| 85 | https://pubmed.ncbi.nlm.nih.gov/29552785 | Journal of B.U.ON. | BAKIS Productions LTD | An image with duplication issues was replaced without notice as described by *Hoya camphorifolia* in February 2021. | https://pubpeer.com/publications/D5C6F71C53559DEF67A6BA4603065E#2 | No |
| 86 | https://pubmed.ncbi.nlm.nih.gov/31424644 | Journal of B.U.ON. | BAKIS Productions LTD | An image with duplication issues was replaced without notice as described by *Hoya camphorifolia* in January 2021. | https://pubpeer.com/publications/DF44180B6D84E2F845DC5378E4A659#3 | No |
| 87 | https://pubmed.ncbi.nlm.nih.gov/31786843 | Journal of B.U.ON. | BAKIS Productions LTD | An image with duplication issues was replaced without notice as described by *Hoya camphorifolia* in January 2021. | https://pubpeer.com/publications/E45369273FA73701DF5052005392BC#2 | No |
| 88 | https://pubmed.ncbi.nlm.nih.gov/31983118 | Journal of B.U.ON. | BAKIS Productions LTD | An image with duplication issues was replaced without notice as described by *Hoya camphorifolia* in February 2021. | https://pubpeer.com/publications/E62858C0783B601AEBFA3EE004CDC5#3 | No, the publication has been reverted to its original form. This has been documented by *Hoya camphorifolia* in June 2022. The stealth correction was present for 15 months at most. |
| 89 | https://pubmed.ncbi.nlm.nih.gov/33277856 | Journal of B.U.ON. | BAKIS Productions LTD | An image with duplication issues was replaced without notice as described by *Hoya camphorifolia* in January 2021. | https://pubpeer.com/publications/E8553755431BEAE89E437DE3816A28#3 | No, the publication has been reverted to its original form. This has been documented by *Hoya camphorifolia* in September 2021. The stealth correction was present for 9 months at most.The publication was retracted in July/August 2021. |
| 90 | https://pubmed.ncbi.nlm.nih.gov/31424652 | Journal of B.U.ON. | BAKIS Productions LTD | An image with duplication issues was replaced without notice as described by *Hoya camphorifolia* in January 2021. | https://pubpeer.com/publications/E964C60689E3ACFB105D56452B8CB7#10 | No, the publication has been reverted to its original form. This has been documented by *Hoya camphorifolia* in June 2022. The stealth correction was present for 18 months at most. |
| 91 | https://pubmed.ncbi.nlm.nih.gov/29332340 | Journal of B.U.ON. | BAKIS Productions LTD | An image with duplication issues was replaced without notice as described by *Hoya camphorifolia* in February 2021. | https://pubpeer.com/publications/EACF0B994A98672EB4501717B684AB#5 | No |
| 92 | https://pubmed.ncbi.nlm.nih.gov/32521857 | Journal of B.U.ON. | BAKIS Productions LTD | An image with duplication issues was replaced without notice as described by *Hoya camphorifolia* in January 2021. | https://pubpeer.com/publications/EBEA63CFA47FAA45D7B3F918A443AC#8 | No, the publication has been reverted to its original form. This has been documented by *Hoya camphorifolia* in September 2021. The stealth correction was present for 9 months at most. An expression of concern has been issued in May/June 2021. |
| 93 | https://pubmed.ncbi.nlm.nih.gov/32277665 | Journal of B.U.ON. | BAKIS Productions LTD | An image with duplication issues was replaced without notice as described by *Hoya camphorifolia* in February 2021. | https://pubpeer.com/publications/EBEBE05DE13C7956B26D4C36ADB17D#4 | No |
| 94 | https://pubmed.ncbi.nlm.nih.gov/29552788 | Journal of B.U.ON. | BAKIS Productions LTD | An image with duplication issues was replaced without notice as described by *Hoya camphorifolia* in February 2021. | https://pubpeer.com/publications/EBFD1BC9DF6A2C1B4D2C026853E0B7#4 | No, the publication has been reverted to its original form. This has been documented by *Hoya camphorifolia* in June 2022. The stealth correction was present for 15 months at most. |
| 95 | https://pubmed.ncbi.nlm.nih.gov/30358200 | Journal of B.U.ON. | BAKIS Productions LTD | An image with duplication issues was replaced without notice as described by *Hoya camphorifolia* in February 2021. | https://pubpeer.com/publications/F5ABB1CFF4EA32EE9AC0480BEAD84D#3 | No |
| 96 | https://pubmed.ncbi.nlm.nih.gov/30570866 | Journal of B.U.ON. | BAKIS Productions LTD | An image with duplication issues was replaced without notice as described by *Hoya camphorifolia* in February 2021. | https://pubpeer.com/publications/F8E86F99D2FD06D10FA56511B94C6E#3 | No, the publication has been reverted to its original form. This has been documented by *Hoya camphorifolia* in July 2022. The stealth correction was present for 18 months at most. |
| 97 | https://pubmed.ncbi.nlm.nih.gov/32521880 | Journal of B.U.ON. | BAKIS Productions LTD | An image with duplication issues was replaced without notice as described by *Hoya camphorifolia* in January 2021. | https://pubpeer.com/publications/FC5B189B1E667069E8E61F87EF680D#3 | No |
| 98 | https://doi.org/10.1016/s0254-6272(16)30012-7 | Journal of Traditional Chinese Medicine | Academy Of Traditional Chinese Medicine | An image with duplication issues was replaced without notice as described by *Spilogona narina* in January 2021. | https://pubpeer.com/publications/F6ED9837A5F42512D30ACC01BDD9EB#4 | No |
| 99 | https://doi.org/10.1109/tcst.2004.839556 | IEEE Transactions on Control Systems Technology | Institute of Electrical and Electronics Engineers | An equation was adjusted without notice as described by *Cyrtodactylus cryptus* in August 2023. | https://pubpeer.com/publications/FC89F2B0439F7F5CA11C6DD7DDB24B#1 | No |
| 100 | https://doi.org/10.1109/tie.2010.2048292 | IEEE Transactions on Industrial Electronics | Institute of Electrical and Electronics Engineers | A word was adjusted without notice as described by *Typhlops schwartzi* in September 2023. This changed the meaning of an explanatory sentence. | https://pubpeer.com/publications/0FC5B75130CA096637B9B5DA72E8E0#1 | No |
| 101 | https://doi.org/10.1016/j.ijbiomac.2015.10.005 | International Journal of Biological Macromolecules | Elsevier | An image with duplication issues was replaced without notice as described by *Elisabeth Bik* in August 2022. | https://pubpeer.com/publications/D08EAD65BBD92909D124B2705B7452#5 | No |
| *Changes in the record of editorial process* | | | | | | |
| 102 | https://doi.org/10.1016/j.jclepro.2020.124781 | Journal of Cleaner Production | Elsevier | The name of the handling editor has been changed without notice as described by *Desmococcus antarctica* in August 2024. | https://pubpeer.com/publications/B6D6D44CCE307C39E1878FE2563B32#3 | No |
| 103 | https://doi.org/10.1016/j.gsf.2024.101804 | Geoscience Frontiers | Elsevier | The name of the handling editor has been changed without notice as described by Wu Guangheng on X (@Dr_5GH) and by Desmococcus antarctica and  Leonid Schneider in June 2024 on Pubpeer. | https://pubpeer.com/publications/0D28D8B0D3BBC66F04FFA6AA832E1F#5 | No |
| 104 | https://doi.org/10.3390/genes14061208 | Genes | MDPI | This paper was initially published in a special issue. The paper, authored by a special issue editor, is now de-listed from that special issue and is now part of a 'section'. Described by *Dorothy Bishop* in August 2024. | https://pubpeer.com/publications/E709DA77E2AFA89FA2222039B667E8#3 | No |
| 105 | https://doi.org/10.3390/agriculture11100944 | Agriculture | MDPI | This paper was initially published in a special issue. The paper, authored by a special issue editor, is now de-listed from that special issue and is now part of a 'section'. Described by *Dorothy Bishop* in August 2024. | https://pubpeer.com/publications/25642BBCBD88801564C92326F7D49C#1 | No |
| 106 | https://doi.org/10.3390/agriculture12060869 | Agriculture | MDPI | This paper was initially published in a special issue. The paper, authored by a special issue editor, is now de-listed from that special issue | https://pubpeer.com/publications/50FF92D83DA1D3277C5E21F2B7B731#1 | No |
| | | | | and is now part of a 'section'. Described by *Dorothy Bishop* in August 2024. | | |
| 107 | https://doi.org/10.3390/agriculture12091313 | Agriculture | MDPI | This paper was initially published in a special issue. The paper, authored by a special issue editor, is now de-listed from that special issue and is now part of a 'section'. Described by *Dorothy Bishop* in August 2024. | https://pubpeer.com/publications/AF42B7D4145ECA1D61D36107713A89#1 | No |
| 108 | https://doi.org/10.3390/agriculture12091330 | Agriculture | MDPI | This paper was initially published in a special issue. The paper, authored by a special issue editor, is now de-listed from that special issue and is now part of a 'section'. Described by *Dorothy Bishop* in August 2024. | https://pubpeer.com/publications/B9B3AE279C457F59793C9F93B46D17#1 | No |
| 109 | https://doi.org/10.3390/agriculture12091436 | Agriculture | MDPI | This paper was initially published in a special issue. The paper, authored by a special issue editor, is now de-listed from that special issue and is now part of a 'section'. Described by *Dorothy Bishop* in August 2024. | https://pubpeer.com/publications/135B0D165FF497E2C64673B563FBD4#1 | No |
| 110 | https://doi.org/ | Agriculture | MDPI | This paper was initially published in a special issue. The paper, authored by a special issue editor, is now de-listed from that special issue and is now part of a 'section'. Described by *Rene Aquarius* in August 2024. | https://pubpeer.com/publications/19A2A53D57EAF9A91DAC2BFFD54307#1 | No |
| 111 | https://doi.org/10.3390/agriculture13010115 | Agriculture | MDPI | This paper was initially published in a special issue. The paper, authored by a special issue editor, is now de-listed from that special issue and is now part of a 'section'. Described by *Rene Aquarius* in August 2024 | https://pubpeer.com/publications/D76D00823A3F65083CEAFEC099A276#1 | No |
| 112 | https://doi.org/10.3390/agriculture13010139 | Agriculture | MDPI | This paper was initially published in a special issue. The paper, authored by a special issue editor, is now de-listed from that special issue and is now part of a 'section'. Described by *Rene Aquarius* in August 2024 | https://pubpeer.com/publications/102D3D0DA487C6C317050A6BEBCA55#1 | No |
| 113 | https://doi.org/10.3390/agriculture13010176 | Agriculture | MDPI | This paper was initially published in a special issue. The paper, authored by a special issue editor, is now de-listed from that special issue and is now part of a 'section'. Described by *Rene Aquarius* in August 2024 | https://pubpeer.com/publications/7B7E73BBDF2CEED6D2E189E441454C#1 | No |
| 114 | https://doi.org/10.3390/agriculture13010224 | Agriculture | MDPI | This paper was initially published in a special issue. The paper, authored by a special issue editor, is now de-listed from that special issue | https://pubpeer.com/publications/3999EBE091B920456F2AF2A1070179#1 | No |
| | | | | and is now part of a 'section'. Described by *Rene Aquarius* in August 2024 | | |
| 115 | https://doi.org/10.3390/agriculture13020412 | Agriculture | MDPI | This paper was initially published in a special issue. The paper, authored by a special issue editor, is now de-listed from that special issue and is now part of a 'section'. Described by *Rene Aquarius* in August 2024 | https://pubpeer.com/publications/A9BE33AEFEF28333A223170047670E#1 | No |
| 116 | https://doi.org/10.3390/agriculture13020436 | Agriculture | MDPI | This paper was initially published in a special issue. The paper, authored by a special issue editor, is now de-listed from that special issue and is now part of a 'section'. Described by *Rene Aquarius* in August 2024 | https://pubpeer.com/publications/AD3E188FE33E948971C5217A7BE2E5#1 | No |
| 117 | https://doi.org/10.3390/agriculture13020446 | Agriculture | MDPI | This paper was initially published in a special issue. The paper, authored by a special issue editor, is now de-listed from that special issue and is now part of a 'section'. Described by *Rene Aquarius* in August 2024 | https://pubpeer.com/publications/C6C778C32BF035775F2A2AD3AE1460#1 | No |
| 118 | https://doi.org/10.3390/agriculture13030645 | Agriculture | MDPI | This paper was initially published in a special issue. The paper, authored by a special issue editor, is now de-listed from that special issue and is now part of a 'section'. Described by *Rene Aquarius* in August 2024 | https://pubpeer.com/publications/9E8C193C5E9912CBE9360ED6A7E380#1 | No |
| 119 | https://doi.org/10.3390/agriculture13040765 | Agriculture | MDPI | This paper was initially published in a special issue. The paper, authored by a special issue editor, is now de-listed from that special issue and is now part of a 'section'. Described by *Rene Aquarius* in August 2024 | https://pubpeer.com/publications/461E2DA4A726413A5327AEDA9286AD#1 | No |
| 120 | https://doi.org/10.3390/agriculture13040839 | Agriculture | MDPI | This paper was initially published in a special issue. The paper, authored by a special issue editor, is now de-listed from that special issue and is now part of a 'section'. Described by *Rene Aquarius* in August 2024 | https://pubpeer.com/publications/E3E3A78EC80B2695D9022BC3E7C3E7#1 | No |
| 121 | https://doi.org/10.3390/agriculture13040899 | Agriculture | MDPI | This paper was initially published in a special issue. The paper, authored by a special issue editor, is now de-listed from that special issue and is now part of a 'section'. Described by *Rene Aquarius* in August 2024 | https://pubpeer.com/publications/A79D7F4B5641F29DD841EB97D4C215#1 | No |
| 122 | https://doi.org/10.3390/agriculture13050975 | Agriculture | MDPI | This paper was initially published in a special issue. The paper, authored by a special issue editor, is now de-listed from that special issue | https://pubpeer.com/publications/26BE85801F38389E8143345324862A#1 | No |
| | | | | and is now part of a 'section'. Described by *Rene Aquarius* in August 2024 | | |
| 123 | https://doi.org/10.3390/agriculture13050997 | Agriculture | MDPI | This paper was initially published in a special issue. The paper, authored by a special issue editor, is now de-listed from that special issue and is now part of a 'section'. Described by *Rene Aquarius* in August 2024 | https://pubpeer.com/publications/2EDD9F8C417DC107A322BCA85D986A#1 | No |
| 124 | https://doi.org/10.3390/agriculture13061142 | Agriculture | MDPI | This paper was initially published in a special issue. The paper, authored by a special issue editor, is now de-listed from that special issue and is now part of a 'section'. Described by *Rene Aquarius* in August 2024 | https://pubpeer.com/publications/686E05B66FB07D83573177279AFCEE#1 | No |
| 125 | https://doi.org/10.3390/agriculture13071338 | Agriculture | MDPI | This paper was initially published in a special issue. The paper, authored by a special issue editor, is now de-listed from that special issue and is now part of a 'section'. Described by *Rene Aquarius* in August 2024 | https://pubpeer.com/publications/77DB9237BEB3E9204741EB8C6B10E0#1 | No |
| 126 | https://doi.org/10.3390/agriculture13081505 | Agriculture | MDPI | This paper was initially published in a special issue. The paper, authored by a special issue editor, is now de-listed from that special issue and is now part of a 'section'. Described by *Rene Aquarius* in August 2024 | https://pubpeer.com/publications/7584D0548C404360E4EED4C6EF364B#1 | No |
| 127 | https://doi.org/10.3390/agriculture13081522 | Agriculture | MDPI | This paper was initially published in a special issue. The paper, authored by a special issue editor, is now de-listed from that special issue and is now part of a 'section'. Described by *Rene Aquarius* in August 2024 | https://pubpeer.com/publications/6F90EE2A125D8BBB48DFE87DF1666B#1 | No |
| 128 | https://doi.org/10.3390/agriculture13081514 | Agriculture | MDPI | This paper was initially published in a special issue. The paper, authored by a special issue editor, is now de-listed from that special issue and is now part of a 'section'. Described by *Rene Aquarius* in August 2024 | https://pubpeer.com/publications/CAC701600741AFE212041E26BEBE1D#1 | No |
| | | | | *Changes in additional information* | | |
| 129 | https://doi.org/10.1016/j.envint.2023.108177 | Environmental International | Elsevier | Ethics statement has been added after publication without notice as described by *Alburnus neretvae* in August 2024. | https://pubpeer.com/publications/A79BA92C456051E62C1FAB00E2B077#4 | No |
| 130 | https://doi.org/10.1016/j.heliyon.2024.e31945 | Heliyon | Elsevier | Consent for publication statement has been changed after publication without notice as described by *Rene Aquarius* in August 2024. | https://pubpeer.com/publications/A0FD169A0FB7CC4C4EDCBDD070CE26#2 | No |
| 131 | https://doi.org/10.1016/j.rsma.2024.103469 | Regional Studies in Marine Science | Elsevier | Final sentence of acknowledgements was changed after publication without notice as described by *Thomas Kesteman* in May 2024. | https://pubpeer.com/publications/C52322D1F5468501DF08A3BA6DBE48#1 | No |